\documentclass{elsart}
\usepackage[dvips]{graphics}
\begin{document}

\parindent=1em
\runauthor{Cicero, Caesar and Vergil}

\begin{frontmatter}   

\begin{flushright}
YGHP-03-31
\end{flushright}

\title{Improved gauge action on an anisotropic lattice II\\
- Anisotropy parameter in the medium coupling region -
}

\author[yamagata]{S.~Sakai}
and
\author[hiro]{A.~Nakamura} 
\address[yamagata]{Faculty of Education, Yamagata University, 
Yamagata 990-8560, Japan}
\address[hiro]{RIISE, Hiroshima University, Higashi-Hiroshima 739-8521, Japan}

\begin{abstract}

The quantum correction of the anisotropy parameter, $\eta$,
is calculated for $\xi=2$ and 3 in the $\beta$ region 
where numerical simulations such as hadron spectroscopy are 
currently carried out,
for the improved actions composed of plaquette and
rectangular 6-link loops.  
The $\beta$ dependences of $\eta$ for the renormalization group improved
actions are quite different from those of the standard and Symanzik actions.
In Iwasaki and DBW2 actions,
$\eta$ stays almost constant in a wide range of $\beta$, which also 
differs from the one-loop perturbative result,
while in the case of 
Symanzik action, it increases as $\beta$ decreases, which is
qualitatively similar to the perturbative result, but the slope is steeper.
In the calculation of the $\eta$
parameter close to and in the confined phase, we have
applied the link integration method to suppress the fluctuation of the 
gauge fields. Some technical details are summarized.

\end{abstract}
\end{frontmatter}


\section{Introduction}
Anisotropic lattices, with temporal lattice spacing smaller than the
spatial one, provide an effective method of precise 
Monte Carlo calculations of, for example, heavy quark systems, glueball masses, 
finite temperature properties of the QCD.
Properties of the anisotropic lattice with the standard plaquette action 
have been studied by several groups\cite{karsch,klassen,eta_bielefeld}.

On the other hand, improved actions are proposed to obtain numerical
results close to the continuum limit on relatively coarse lattices.
Therefore it is worth studying the anisotropic properties of the
improved actions.

In the previous paper, we studied the anisotropic lattice for
a class of improved actions in the weak coupling region, mainly using
the perturbative method\cite{sakai}. 
The improved actions we considered are
written in terms of the plaquette and 6-link rectangular loops as
\begin{equation}
 S \propto \sum \left[C_0 P(1 \times 1)_{\mu \nu}
 + C_1  P(1 \times 2)_{\mu \nu}   \right] ,
\label{impaction}
\end{equation}
where $C_0$ and $C_1$ satisfy the relation $C_0 + 8C_1=1$. 
The improved actions frequently used in the simulations
correspond to the following parameters:
$C_1=-1/8$  (Symanzik's improved action\cite{symanzik}), 
$C_1=-0.331$ (Iwasaki's improved action\cite{iwasaki}), 
and $C_1=-1.4088$ (QCDTARO's DBW2 action\cite{dbw2}).

For these types of actions, we can formulate the anisotropic lattice 
in the same way as the standard plaquette action,
\begin{equation}
  S_g = \beta_{\xi}(\frac{1}{\xi_{B}}
   \sum_x\sum_{i>j}P_{ij}+
      \xi_{B}
   \sum_x\sum_{i\neq4}P_{4i}) ,
\label{anisoaction}
\end{equation}
where $\beta_{\xi}=\sqrt{\beta_{\sigma} \beta_{\tau}}$, and
$\xi_B$ is a bare anisotropic parameter which controls the anisotropy
in the space and time directions.
The anisotropy is defined as the ratio of the lattice spacing in the
spatial ($a_{\sigma}$) and the temporal ($a_{\tau}$) directions,
 $\xi_R=a_{\sigma}/a_{\tau}$. 

Due to quantum correction, $\xi_R$ is not equal to $\xi_B$.
Therefore it is essential to know their relationship
before large scale simulations on the anisotropic lattice with improved actions
are carried out. 

The effects of quantum correction have been studied using the
parameter $\eta$ defined by
\begin{equation}
 \eta=\frac{\xi_R}{\xi_B} .
\label{eta}
\end{equation}
In the weak coupling region, it is calculated
perturbatively\cite{karsch},\cite{sakai}.
The one-loop perturbative results have been very
impressive in the sense that as $-C_1$ increases, qualitative change
is observed in the behavior of $\eta$
as a function of $\beta$.
If it is parametrized as
\begin{equation}
\eta(\xi,\beta,C_1)=1+\frac{N_c}{\beta}\eta_1(\xi,C_1) ,
\label{pertetadef}
\end{equation}
the coefficient $\eta_1$ decreases as $-C_1$ increases,
and at around $C_1 \sim -0.18$, 
it reaches to zero and then becomes negative. 
Therefore $\eta(\beta)$ of Iwasaki action and DBD2
action decreases as $\beta$ decreases, while in the case of the standard
and Symanzik actions, it increases. 
Namely, they have opposite $\beta$ dependences.

The natural question is what would be the behavior of $\eta$ in the
smaller $\beta$ region, where the perturbative calculation breaks down.
In this work, we will focus on the $\eta$ parameter at $\xi_{R}=2$ and
$3$, and calculate it
in the intermediate $\beta$ region where most current
numerical simulations are carried out.

In Section 2, we discuss appropriate regions of $\beta$ 
for the 
improved actions to evaluate $\eta$, 
and explain some details of the calculation:
matching of the lattice potential in the spatial and temporal
directions, and the method of eliminating the effects of 
self-energy terms of lattice potential on an anisotropic lattice.
For the standard action, the effect of the self-energy terms on $\eta$
has been studied by the Bielefeld group \cite{eta_bielefeld}. 
They reported that the effects on $\eta$ are small about $1\%$
throughout the parameter range that they studied.
We discuss the effects for improved actions and
show, in Section 3, that they are not large.

In Section 3, $\eta$ for each improved action is presented. 
The $\eta$ behavior in the intermediate $\beta$ region 
is quite different for each improved action.
Iwasaki and DBW2 actions are qualitatively different
in the one-loop perturbative calculation.

Section 4 is devoted to the discussion and conclusion.
It is found that for Iwasaki action, $\eta$ is close to unity in the
region $\beta \geq 2.5$,
indicating that the detailed calibration 
of $\xi_B$ for a given $\xi_R$ is not important 
unless a very precise simulation is carried out.
For DBW2 action, $\eta$ remains essentially constant,
indicating that rough calibration 
gives a reasonable estimation of $\eta$. For Symanzik action some detailed 
calculation of $\eta$ is necessary, as in the case of standard action.

For the calculation of the $\eta$ parameter,
measurements of the large Wilson loops are required. 
Large Wilson loops suffer from huge 
fluctuation of the gauge fields, particularly in the confined phase.
To suppress the fluctuation, the link integration method has been
proposed \cite{parisi,brower,forcrand}.
In this work, we have applied the link integration method
in small $\beta$ regions.
Here, it is very important to choose
an adequate radius (optimal radius) of integration in the complex plane.
For the standard action, 
the optimal radius was studied by the Bielefeld Group \cite{eta_bielefeld}. 
In appendix A, we describe the optimal radii for Symanzik and
Iwasaki actions.

\section{Calculation of the anisotropy quantum correction, $\eta$}

\subsection{Region of coupling constant to be studied}

In this work,
we calculate the $\eta$ parameter in the regions of $\beta$ where
most numerical calculations are currently carried out. 
In the case of standard action, hadron
spectroscopy in the quench approximation has been reported for 
$5.7 < \beta < 6.2$ \cite{yoshie}. 
In these coupling constant regions,
the light hadron masses are reproduced within $10\%$ accuracy, 
which may be the limit of the quench approximation. 
Therefore, we calculate the $\eta$ parameter around
these lattice spacings for the improved actions.

In order to estimate the lattice spacing for the improved actions, we 
use the critical $\beta$ ($\beta_{Crit}$) of finite temperature
transition. 
For standard action,  $\beta=6.05$ corresponds to the
finite temperature transition point of the $N_T=8$ 
lattice\cite{qcdtaro_compilation}. Therefore 
we estimate $\beta_{Crit}$ at $N_T=8$ for the improved actions.
For tree level improved Symanzik action, $\beta_{Crit}$'s
are reported for $N_{T}=3,4,5$ and 6\cite{cella}, and for 
Iwasaki action, they have been calculated at  
$N_{T}=4, 6$ by the Tsukuba group\cite{kaneko} and at $N_{T}=8$, by
the Yamagata-Hiroshima collaboration\cite{sakai96}. 
For DBW2 action, they are reported by the QCDTARO collaboration\cite{dbw2} 
for $N_{T}=3, 4$ and 6.

$\beta_{Crit}$ at $N_{T}=8$ is estimated using the 
two loop asymptotic scaling relation for lattice spacing,
\begin{equation}
   a= \frac{1}{\Lambda} (\frac{6b_0}{\beta})^{-c}\exp(-\frac{\beta}{12b_0}),
\label{ainv}
\end{equation}
where $a$ is the lattice spacing and $c=b_1/(2 b_0^2$), 
with $b_0=11/(4\pi)^2$ and $b_1=102/(4\pi)^4$.
We applied two methods for the estimation of $\beta_{Crit}$ at $N_T=8$.
In method 1, we use $\beta_{Crit}$ at $N_T=6$ of the same action 
to estimate $\beta_{Crit}(N_T=8)$ using 
$T_c=1/(aN_T)$.
In method 2, we use the $\beta_{Crit} (N_T=8)$ of the standard action
to evaluate $\beta_{Crit}(N_T=8)$ of the other action
with the $\Lambda$ parameter\cite{sakai,lambda}.
They are summarized in the table \ref{table1}.\\
\renewcommand{\arraystretch}{1.0}
\begin{table}[h]
\begin{center}
\caption{Estimation of $\beta_{Crit}$ at $N_{T}=8$ for various actions}
\label{table1}
\begin{tabular}{|c|c|c|c|c|c|c|}
     \hline
     \multicolumn{1}{|c|}{ } &
     \multicolumn{1}{|c|}{$\beta_{Crit}(data)$} &
     \multicolumn{1}{|c|}{Method 1}&
     \multicolumn{1}{|c|}{Method 2}&
     \multicolumn{1}{|c|}{Minimum $\beta$}\\
     \hline
        $Standard$     &6.05($N_{T}=8$)   &      &              & \\
     \hline
        $Symanzik$     &4.31($N_{T}=6$)  &4.57   &4.56          &4.5\\
     \hline
        $Iwasaki $     &2.52($N_{T}=6$)  &2.78   &2.32          &2.5\\
     \hline
        $DBW2$         &0.936($N_{T}=6$) &1.28   &-             &1.1\\ 
     \hline
\end{tabular}
\end{center}
\end{table}
\indent  
In the case of Symanzik action,
the estimations of $\beta_{Crit}$ by the two methods coincide with each
other. For Iwasaki action, some discrepancy is observed between 
the two estimations. 
The first method gives a closer result to that of Ref.\cite{sakai96}, 
in which $\beta_{Crit}= 2.73 \sim 2.75 $.

For DBW2 action, $\beta_{Crit}$ estimated using the
$\Lambda$ ratio becomes negative. 
In this $\beta$ region, the deviation from the
perturbative scaling relation is quite large for this action.
Therefore, for the estimation of $\beta_{Crti} (N_T=8)$, we 
plot $\beta_{Crit}$ at $N_t=3,4$ and 6 and simply
extrapolate it.
It becomes about $\beta=1.1$ with large ambiguity.
Therefore we will study the action until $\beta=1.0$.

In the table \ref{table1}, we show also the minimum of $\beta$, for
which we calculate the $\eta$ parameters.

\subsection{Subtraction of the self-energy contribution from the
lattice potential}

The renormalized anisotropy $\xi_{R}$ is defined as the ratio of the
lattice spacings in the spatial and temporal directions,
$\xi_{R}=a_{\sigma}/a_{\tau}$.
In the quench approximation, the lattice potential has been used as
the probe of lattice spacing, and is defined as the ratio of
Wilson loops,
\begin{equation}
V(p,r)=\log(\displaystyle{\frac{W(p,r)}{W(p+1,r)}}).
\label{latpot}
\end{equation}
The lattice potentials in the spatial ($V_{s}$) and  temporal
directions ($V_{t}$) are
determined by the Wilson loops in the space-space plane and 
the space-time plane, respectively.
The lattice potential defined in Eq.\ref{latpot} will become
independent of the position $p$ when $p$ becomes large
where the lattice artifact disappears. 
In the following, we will discard $p$ 
in the artifact-free region, but write it explicitly when we
discuss its effect.

The matching of the potential in the spatial and temporal 
directions \cite{klassen,nakamura} has been used for the
determination of $\xi_{R}$ as
\begin{equation} 
  V_{s}(\xi_{B},r)=V_{t}(\xi_{B},t=\xi_{R} \times r) .
\label{match}
\end{equation}  
In the calculation of $\eta$, we fix the renormalized anisotropy
$\xi_R$, and then search for the point of $\xi_B$ where the Eq.\ref{match} is
satisfied \cite{klassen}.
Using these $\xi_B$ and $\xi_R$ values, the $\eta$ parameter is calculated.

The lattice potential defined by Eq. \ref{latpot}
suffers from the self-energy term.
In this article, we assume the simplest parameterization for it:
\begin{equation}
  V_{s}(\xi_{B},r)=V_{s}^{0}(\xi_{B})+V_{s}^{L}(\xi_{B},r),
\label{nosubpot}
\end{equation}  
where $V_{s}^{L}$ is the lattice potential free of self-energy
contributions.
The temporal potential, $V_{t}(\xi_{B},t)$, is treated similarly. 
On an anisotropic lattice, 
$V_{s}^{0}(\xi_{B})$ and $V_{t}^{0}(\xi_{B})$ may be different from each 
other due to the anisotropy. 
Therefore the matching of the potential should be applied for 
the self-energy-free parts, $V_s^{L}$ and $V_t^{L}$.

In order to eliminate the effect of the self-energy term $V^{0}$, 
we define the subtracted potential:
\begin{equation}
 V_{s}^{Sub}(\xi_{B},r,r_0)=V_{s}(\xi_{B},r)-V_{s}(\xi_{B},r_{0})
           =V_{s}^{L}(\xi_{B},r)-V_{s}^{L}(\xi_{B},r_{0}) .
\label{potsub}
\end{equation}
$V_{t}^{Sub}$ is defined in a similar manner.

The subtraction points $r_{0}$ and $t_{0}$ are chosen to
satisfy $t_{0}=\xi_{R} r_{0}$, and the matching of the
potential,
$V_{t}^{L}(t_{0}=\xi_{R}r_{0})=V_{s}^{L}(r_{0})$, should also be
satisfied there.
Namely, at $r_0$, the lattice potential should be free of the lattice
artifacts. This condition is satisfied if $r_0$ is large. 
At large $r_0$, however, the fluctuation of the potential increases, 
and simulations with high statistics on a larger lattice are required. 
Therefore, $r_0$ should be chosen to be as small as possible, 
where the lattice artifacts will be sufficiently small.

\subsection{Matching method}

As an example, we will show details of the determination of $\eta$ 
in the case of Iwasaki action at
$\beta=4.5$, $\xi_R=2$ on the $12^3 \times 24$ lattice.

Let us start with the determination of the subtraction point $r_0$.
In order to reduce statistical error,
small $r_{0}$ is preferable. 
In the small $r_{0}$ region, however,
systematic error due to lattice artifacts becomes large. 
The optimal choice of $r_{0}$ requires careful study by trial and
error.

\begin{figure}
\begin{center}
\scalebox{0.65}{ { \includegraphics{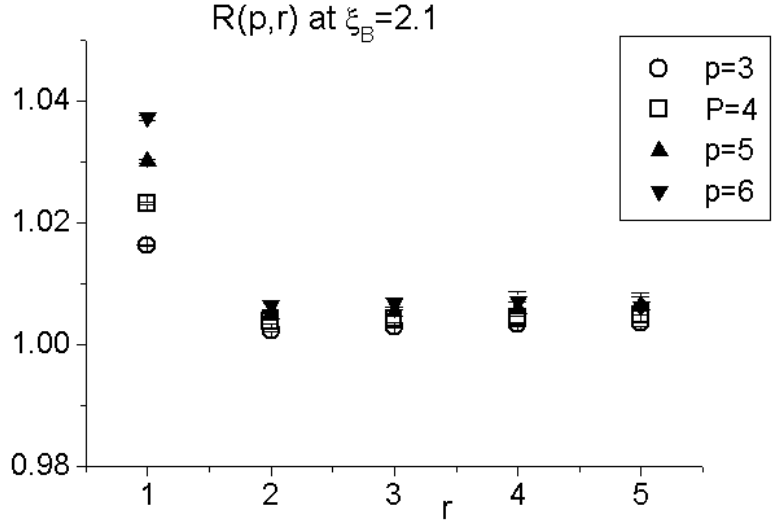} } }
\caption{Ratios given by Eq.\ref{rate_nosub}.}  
\label{rate_p}
\end{center}
\end{figure}

First we calculate the ratio
\begin{equation}
R(\xi_{B},p,r)=\frac{V_{s}(\xi_{B},p,r)}
{V_{t}(\xi_{B},p,t=\xi_{R}\times r)} ,
\label{rate_nosub}
\end{equation}
where $V_{s}$ and $V_{t}$ mean the lattice potential 
in the space and time directions, respectively, 
and include the self-energy contributions.  
Our results are displayed in the Fig. \ref{rate_p}.
As $r$ increases, the ratio
$R(p,r)$ approaches an asymptotic value.
It is seen that at $r=1$, the deviation from the asymptotic value is
rather large, which may be due to lattice artifacts.
Therefore, we first choose $r_0=3$, 
and calculate the subtracted potentials of Eq.\ref{potsub}, and
then use them in Eq.\ref{rate_nosub} to obtain ratio $R$.

\begin{figure}
\begin{center}
\scalebox{0.65}{ { \includegraphics{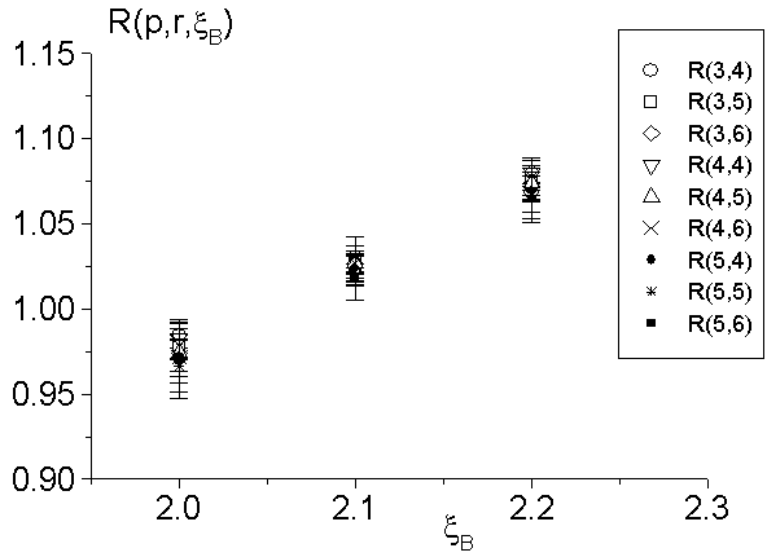} } }
\caption{$\xi_{B}$ dependences of the ratios for the subtracted potentials.}  
\label{rate_xi}
\end{center}
\end{figure}
\begin{figure}
\begin{center}
\scalebox{0.65}{ { \includegraphics{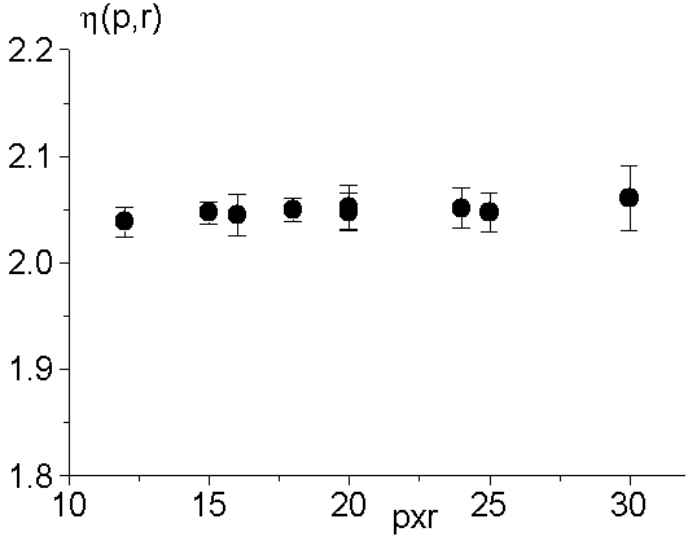} } }
\caption{$\eta=\xi_R/\xi_B$ at each $p$ and $r$, where $\xi_R$ ($\xi_B$)
stands for the renormalized (bare) anisotropy.}  
\label{eta_sampl}
\end{center}
\end{figure}

   
The results for $\xi_B=2.0$, 2.1 and 2.2 are shown in Fig.\ref{rate_xi}. 
The ratios are shown individually for each $p$ and $r$.  We proceed to
look for the point where the ratios satisfy the relation
$R(p,r,\xi_{B})=1$. 
We fit the three points by the second-order polynomial of
$\xi_B$ and find the solution
\begin{equation}
  R(p,r,\xi_{B})= c_0 + c_1 \xi_B + c_2 \xi_B^2 =1.
\label{sol_xi}
\end{equation}  
The coefficients $c_0$, $c_1$ and $c_2$ are determined by the three
data points of $R(p,r,\xi_B)$. 
Using the solution of Eq.\ref{sol_xi},
the ratios $\eta=\xi_R/\xi_B$ are determined  for each $p$ and $r$
and are shown in Fig.\ref{eta_sampl}.

In order to avoid the lattice artifact, we choose the data with
$p \ge 3$ and $r \ge 4$. In this region, $\eta(p,r)$ is almost
independent of $p$ and $r$. 
$\eta$ at this $\beta$ is determined as the average of the data. 
The errors are estimated
by the jackknife method; the data after thermalization is grouped into
10 blocks and they are used as independent data.
In this way the result becomes $\eta=0.9766 \pm
0.0044$ when $r_0=3$.

The same analysis is carried out by taking
$r_{0}=2$. The result becomes $\eta=0.9764 \pm 0.0039$. 
The result changes little in this case. 
However, if we choose $r_0=1$, the
results differ significantly from those of $r_0=$2 and 3.
Analyses are carried out at other values of $\beta$ and $\xi$. 
There are several cases in which a 
slight difference is observed between $r_0=2$ and $r_0=3$.
Therefore in the case of Iwasaki action, we choose
$r_0=3$ for all values of $\beta$ and $\xi_B$ in this work.

We carried out the same studies for Symanzik and DBW2 actions.
In these cases, the subtraction point becomes
$r_0=4$. This indicates that lattice artifacts are larger for 
these actions.

\section{Result for the quantum correction of the anisotropy,
$\eta$}

\subsection{Simulation parameters, numerical results and self-energy
  contribution}

The simulations are mainly carried out on the $12^3 \times 12 \xi_R$
lattice.
For some values of $\beta$ and $\xi$, a larger
$16^3 \times 16 \xi_R$ lattice is employed in order to study the size
dependence.
It is found that the lattice size effect is small.

The gauge configurations are generated by the heat bath
method with overrelaxation \cite{woch,creutz2}.
Typical numbers of the Monte Carlo (MC) data for the calculation of
$R(\xi_B,p,r)$ are a few tens of thousand
after thermalization of about $10^4$ MC sweeps.  
However, as $\beta$ decreases and approaches the finite transition point 
or goes into the confined phase, both the necessary number of MC data
and the number of thermalization sweeps increase.
For the calculation of $\eta (\xi=2)$ at
$\beta=2.5$ for Iwasaki action, we used $1.5\times10^6$ data after
thermalization of $3.5\times 10^5$ MC sweeps. 

In order to suppress the fluctuation of the gauge fields in the
calculation of large Wilson loops, 
we applied the link integration method \cite{brower,forcrand}. 
It is used for the calculation of the lattice potentials 
at $\beta=2.5$ and 2.56 for
Iwasaki action and $\beta=4.5$ for Symanzik action.
Technical details will be presented in the appendix. 
Here we only notice that, in the case of improved actions, 
the effect of the link integration is
reduced due to the presence of the rectangular 6-link loops.

Our results of $\eta$ are
summarized in the tables \ref{Syman_eta_2} to \ref{DBW2_eta_3}.
In order to show the effects of self energy terms in the
lattice potential,
we have presented the results for $\eta$, which are obtained
without subtracting the self energy terms in the 
$\eta^{Nosub}$ column in these tables. It is found that the difference
between them is less than $\sim 1\%$ for Symanzik and Iwasaki
actions. This is consistent with the result for the standard action
obtained by Bielefeld group\cite{eta_bielefeld}. 
However, for DBW2 action, the difference increases. 
It amounts to a few percent but is still small. 
Therefore, except for the case of the simulation with a few
percent accuracy, it is safe to use $\eta^{Nosub}$, which has been
reported at the XVIIth International Symposium on Lattice Field Theory at
Pisa (Lattice '99) \cite{sakai2}.

\renewcommand{\arraystretch}{1.0}
\begin{table}[h]
\begin{center}
\caption{ $\eta \equiv \xi_R/\xi_B$ for Symanzik action at $\xi_{R}=2$,
where $\xi_R$ and $\xi_B$ are the renormalized and bare anisotropies,
respectively.  
At $\beta=8.0$ and 4.5  the simulation is carried out on 
$16^3 \times 32$ lattice to study the size dependences. 
They are shown in the line with the symbol $*$.
  }
\label{Syman_eta_2}
\begin{center}
     \vspace {0.5cm}
     Symanzik Action at  $\xi_R=2$ 
     \vspace {0.5cm}
\end{center}
\begin{tabular}{|c|c|c|c|c|c|c|}
     \hline
     \multicolumn{1}{|c|}{$\beta$} &
     \multicolumn{1}{|c|}{$\eta$}&
     \multicolumn{1}{|c|}{$\eta^{NoSub}$}\\

     \hline
        $10.0$  &1.0227 $\pm$ 0.0097 &1.0271 $\pm$ 0.0031\\
     \hline
        $8.0$   &1.0393 $\pm$ 0.0191 &1.0391 $\pm$ 0.0020\\
     \hline
        $6.0$   &1.0381 $\pm$ 0.0097 &1.0500 $\pm$ 0.0029\\
     \hline
        $4.5$   &1.0980 $\pm$ 0.0255 &1.1011 $\pm$ 0.0021\\
     \hline
     \hline
        $8.0^*$ &1.0284 $\pm$ 0.0021 &1.0232 $\pm$ 0.0039\\
     \hline
        $4.5^*$ &1.1095 $\pm$ 0.0122 &1.1040 $\pm$ 0.0062\\
     \hline
\end{tabular}
\end{center}
\vspace{0.5cm}
\end{table}  
\renewcommand{\arraystretch}{1.0}
\begin{table}[h]
\begin{center}
\caption{ $\eta$ as a function of $\beta$ for Symanzik action at
$\xi_{R}=3$.}
\label{Symanzik_eta_3}
\begin{center}
     \vspace {0.5cm}
     Symanzik Action at  $\xi_R=3$ 
     \vspace {0.5cm}
\end{center}
\begin{tabular}{|c|c|c|c|c|c|c|}
     \hline
     \multicolumn{1}{|c|}{$\beta$} &
     \multicolumn{1}{|c|}{$\eta$}&
     \multicolumn{1}{|c|}{$\eta^{NoSub}$}\\

     \hline
        $10.0$  &1.0341 $\pm$ 0.0146 &1.0426 $\pm$ 0.0058\\
     \hline
        $8.0$   &1.0260 $\pm$ 0.0150 &1.0361 $\pm$ 0.0042\\
     \hline
        $6.0$   &1.0520 $\pm$ 0.0200 &1.0667 $\pm$ 0.0015\\
     \hline
        $4.5$   &1.1482 $\pm$ 0.0317 &1.1331 $\pm$ 0.0064\\
     \hline
\end{tabular}
\end{center}
\vspace{0.5cm}
\end{table}  
\renewcommand{\arraystretch}{1.0}
\begin{table}[h]
\begin{center}
\caption{ $\eta$  for Iwasaki action at
$\xi_{R}=2$. The data for $\beta=3.5$ with * are calculated on the 
$16^3 \times 32$ lattice to study the size dependences.}
\label{Iwasaki_eta_2}
\begin{center}
     \vspace {0.5cm}
     Iwasaki Action at  $\xi_R=2$ 
     \vspace {0.5cm}
\end{center}
\begin{tabular}{|c|c|c|c|c|c|c|}
     \hline
     \multicolumn{1}{|c|}{$\beta$} &
     \multicolumn{1}{|c|}{$\eta$}&
     \multicolumn{1}{|c|}{$\eta^{NoSub}$}\\

     \hline
        $10.0$  &0.9811 $\pm$ 0.0030 &0.9742 $\pm$ 0.0033\\
     \hline
        $6.0$   &0.9831 $\pm$ 0.0037 &0.9784 $\pm$ 0.0032\\
     \hline
        $4.5$   &0.9755 $\pm$ 0.0083 &0.9776 $\pm$ 0.0044\\
     \hline
        $4.0$   &0.9806 $\pm$ 0.0074 &0.9782 $\pm$ 0.0039\\
     \hline
        $3.5$   &0.9761 $\pm$ 0.0105 &0.9767 $\pm$ 0.0049\\
     \hline
        $3.05$  &0.9911 $\pm$ 0.0182 &0.9881 $\pm$ 0.0060\\
     \hline
        $2.5$   &0.9998 $\pm$ 0.0045 &0.9837 $\pm$ 0.0074\\
     \hline
     \hline
        $3.5^*$ &0.9803 $\pm$ 0.0070 &0.9891 $\pm$ 0.0036\\
     \hline
\end{tabular}
\end{center}
\vspace{0.5cm}
\end{table}  
\renewcommand{\arraystretch}{1.0}
\begin{table}[h]
\begin{center}
\caption{ $\eta$ for Iwasaki action at $\xi_{R}=3$.  }
\label{Iwasaki_eta_3}
\begin{center}
     \vspace {0.5cm}
     Iwasaki Action at  $\xi_R=3$ 
     \vspace {0.5cm}
\end{center}
\begin{tabular}{|c|c|c|c|c|c|c|}
     \hline
     \multicolumn{1}{|c|}{$\beta$} &
     \multicolumn{1}{|c|}{$\eta$}&
     \multicolumn{1}{|c|}{$\eta^{NoSub}$}\\

     \hline
        $10.0$  &0.9714 $\pm$ 0.0054 &0.9647 $\pm$ 0.0006\\
     \hline
        $6.0$   &0.9569 $\pm$ 0.0041 &0.9554 $\pm$ 0.0026\\
     \hline
        $4.0$   &0.9700 $\pm$ 0.0118 &0.9645 $\pm$ 0.0063\\
     \hline
        $3.5$   &0.9715 $\pm$ 0.0160 &0.9708 $\pm$ 0.0031\\
     \hline
        $3.05$  &0.9725 $\pm$ 0.0120 &0.9776 $\pm$ 0.0037\\
     \hline
        $2.56$  &1.0011 $\pm$ 0.0138 &1.0011 $\pm$ 0.0071\\
     \hline
\end{tabular}
\end{center}
\vspace{0.5cm}
\end{table}  
\renewcommand{\arraystretch}{1.0}
\begin{table}[h]
\begin{center}
\caption{$\eta$ as a function of $\beta$ for DBW2 action at
$\xi_{R}=2$.}
\label{DBW2_eta_2}
\begin{center}
     \vspace{0.5cm}
     DBW2 action at $\xi_R=2$ 
     \vspace {0.5cm}
\end{center}
\begin{tabular}{|c|c|c|c|c|c|c|}
     \hline
     \multicolumn{1}{|c|}{$\beta$} &
     \multicolumn{1}{|c|}{$\eta$}&
     \multicolumn{1}{|c|}{$\eta^{NoSub}$}\\

     \hline
        $2.5$   &0.9084 $\pm$ 0.0090   &0.8626 $\pm$ 0.0025  \\
     \hline
        $1.6$   &0.9011 $\pm$ 0.0082   &0.8616 $\pm$ 0.0018  \\
     \hline
        $1.4$   &0.8917 $\pm$ 0.0122   &0.8623 $\pm$ 0.0024  \\
     \hline
        $1.2$   &0.8882 $\pm$ 0.0115   &0.8673 $\pm$ 0.0032\\
     \hline
        $1.1$   &0.8868 $\pm$ 0.0144   &0.8753 $\pm$ 0.0030\\
     \hline
        $1.0$   &0.8781 $\pm$0.01069   &0.8817 $\pm$ 0.0092\\
     \hline
\end{tabular}
\end{center}
\end{table}  

\renewcommand{\arraystretch}{1.0}
\begin{table}[h]
\begin{center}
\caption{$\eta$ for DBW2 action at $\xi_{R}=3$.}
\label{DBW2_eta_3}
\begin{center}
     \vspace{0.5cm}
     DBW2 action at $\xi_R=3$ 
     \vspace {0.5cm}
\end{center}
\begin{tabular}{|c|c|c|c|c|c|c|}
     \hline
     \multicolumn{1}{|c|}{$\beta$} &
     \multicolumn{1}{|c|}{$\eta$}&
     \multicolumn{1}{|c|}{$\eta^{NoSub}$}\\

     \hline
        $1.4$   &0.8283 $\pm$ 0.0189 &0.8082 $\pm$ 0.0046\\
     \hline
        $1.2$   &0.8157 $\pm$ 0.0252 &0.8070 $\pm$ 0.0055\\
     \hline
        $1.1$   &0.8122 $\pm$ 0.0230 &0.8210 $\pm$ 0.0076\\
     \hline
        $1.0$   &0.8123 $\pm$ 0.0235 &0.8262 $\pm$ 0.0101\\
     \hline
\end{tabular}
\end{center}
\end{table}  
\indent

\subsection{Symanzik action}

\begin{figure}
\begin{center}
\scalebox{0.55}{ { \includegraphics{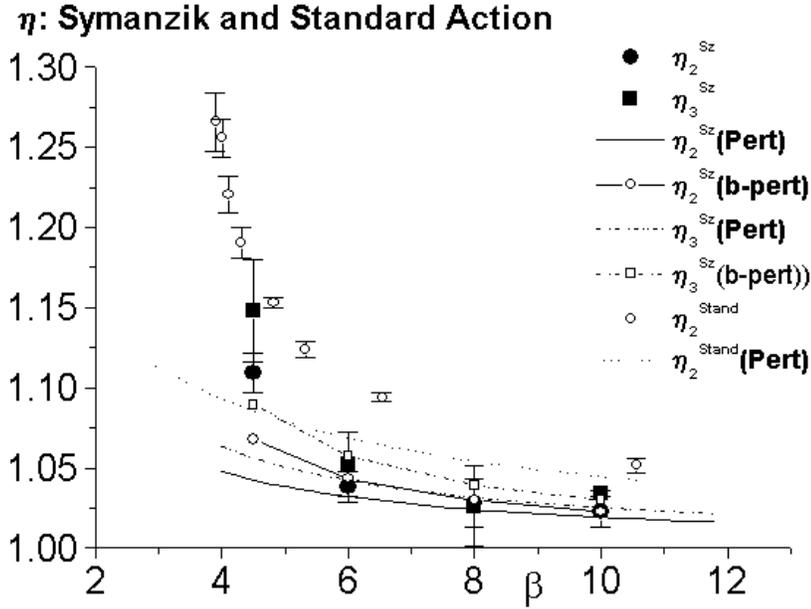} } }
\caption{ $\eta$s for Symanzik ($\eta^{Sz}$) and standard
($\eta^{Stand}$) actions.
For standard action, we show the results obtained by Klassen\cite{klassen}. 
$\beta$ of standard action is shifted by Eq.\ref{ainv}, 
in order to compare the $\eta$ at the same lattice spacing $a$.
The perturbative results are also shown for comparison with
numerical results. The $\eta(b-pert)$ represents the boosted
perturbation result(Eq. \ref{boostpert}).}
\label{eta_sub}
\end{center}
\end{figure}

In Fig.\ref{eta_sub}, we show the values of the $\eta$ parameter 
as a function of $\beta$ for Symanzik action. 
The qualitative behavior of $\eta$ as a function of $\beta$ is
the same in perturbative and numerical results. 
However, the slope of $\eta$ becomes steeper for the numerical
results. \\
\indent
In this case the tadpole improved one-loop perturbation
calculation (boosted perturbation) \cite{petronzio,lepage}
reduces the discrepancy.
It may be to replace $\beta$ in Eq. \ref{pertetadef}
by $\tilde{\beta}=\beta \sqrt{W_s(1,1)W_t(1,1)}$:
\begin{equation}
 \eta(\xi,\beta)=1+
   \frac{N_c}{\beta}\frac{\eta_1(\xi)}{\sqrt{W_s(1,1)W_t(1,1)}}.
\label{boostpert}
\end{equation}
In this formula, since $W_s(1,1)$ and $W_t(1,1)$ decrease as $\beta$
decreases, the $\beta$ dependence of $\eta$ is more enhanced.
The use of Eq.\ref{boostpert} improves the
agreement between perturbative and numerical results a little as 
shown in the Fig. \ref{eta_sub}.

In this figure, we also show the result from the standard plaquette 
action\cite{klassen}.
In order to compare $\eta$ at the same lattice spacing, we have
shifted $\beta$ of standard action to that of Symanzik
action using the asymptotic scaling relation Eq. \ref{ainv}.
Because the two estimations coincide for Symanzik action
for $\beta_{Crit}$ at the $N_T=8$ lattice in Section 2.1, 
in these regions the asymptotic scaling relation Eq.\ref{ainv}
may be satisfied at least approximately for these two actions. 
It is found that the qualitative behaviors are the same, although
the slope becomes more gentle for Symanzik action.

%
%

\subsection{Iwasaki Action}
\begin{figure}
\begin{center}
\scalebox{0.55}{ { \includegraphics{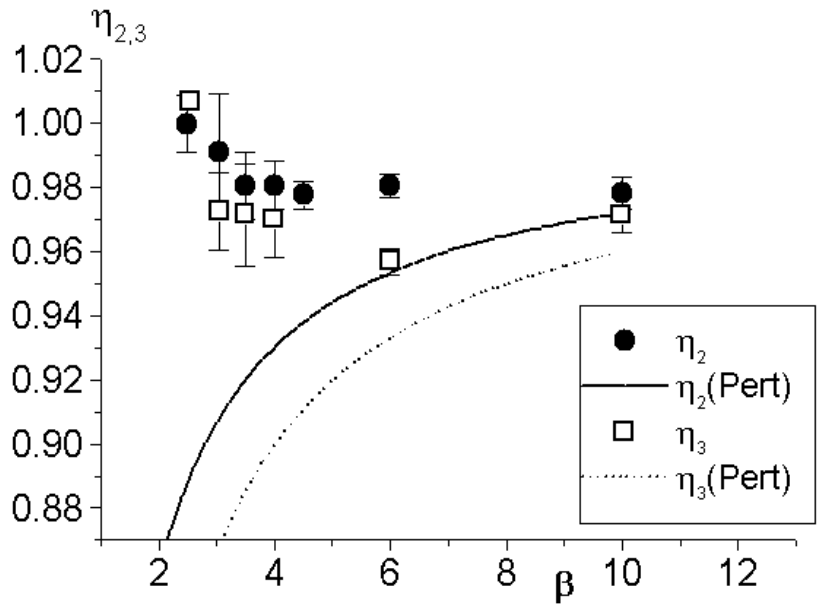} } }
\caption{$\eta$ for Iwasaki action. }
\label{eta_a_Iwa}
\end{center}
\end{figure} 

Results for Iwasaki action are shown in Fig.\ref{eta_a_Iwa}.
$\eta_2$ exhibits a shallow
dip around $\beta \sim 4.5$ and then increases with
decreasing $\beta$. 
$\eta_3$ shows similar behavior, but the position
of the dip moves to $\beta \sim 6.0$. 
Both $\eta_2$ and $\eta_3$ stay close to unity 
in a wide range of $\beta$ for $\beta \geq 2.5$.  
The deviation from unity is more enhanced for $\eta_3$.

In the continuum limit, the $\eta$ parameter approaches one.
The improved action has an $\eta$ value that remains one 
in a wide range of $\beta$; $ 2.5 \le \beta$.
The deviation from unity is within  $4\%$ when $\xi_R=2$, and $3$,
therefore until precision simulation of a few $\%$ accuracy
is required, detailed calibration of $\eta$ is not necessary.
This is a good property for simulations.

For Iwasaki action,
the one-loop perturbative calculation predicts a monotonic
decrease in $\eta$ as $\beta$ decreases\cite{sakai}, 
as shown in Fig.\ref{eta_a_Iwa}. 
The numerical results are qualitatively different from
those of the one-loop perturbation. 

%
%
\subsection{DBW2 Action}

\begin{figure}
\begin{center}
\scalebox{0.75}{ { \includegraphics{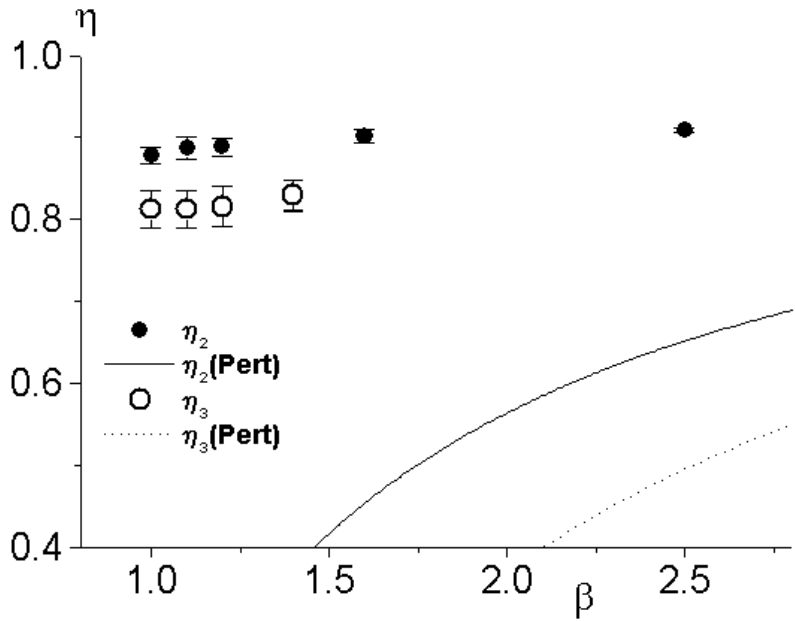} } }
\caption{$\eta$ for DBW2 action.}
\label{eta_2_dbw2}
\end{center}
\end{figure} 

Results for DBW2 action are shown in Fig.\ref{eta_2_dbw2}.
$\eta$ stays almost constant;
$\eta_2 \sim 0.9$ in the range $1.0 \geq \beta \geq 2.5$ 
and $\eta_3 \sim 0.81$ in the range $1.0 \geq \beta \geq 1.4$.
The deviation from unity is not small in this case, but 
the flatness of $\eta$ as a function of $\beta$  
is again a good property for numerical simulations. 
The rough calibration of $\eta$ at a few $\beta$ points
is sufficient to obtain a reasonable estimation of $\xi_B$. 
As in the case of Iwasaki action, the numerical results show 
qualitatively different behavior 
from the one-loop perturbative ones \cite{sakai}
(see Fig.\ref{eta_2_dbw2}). 

\section{Discussion and Conclusions}

In this work, we studied the global structure of $\eta$
as a function of $\beta$ and $C_1$ for the gauge
action given in Eq.\ref{impaction}.
Overall effects of the improved actions on $\eta$ are summarized as
follows. The plaquette term in the action makes 
the $\eta$ parameter increase monotonically as $\beta$ decreases, 
while the rectangular terms with $C_1 <0$ make $\eta$ decrease.

At $C_1=-1/8$, the effects of rectangular loops are not so large, 
and the slope of $\eta$ is smaller than that in the case
of standard action. 
As a result, at the same lattice spacing, 
the effects of the quantum correction are reduced in Symanzik action.
In Symanzik and standard actions, the $\eta$ dependence on $\beta$ is
qualitatively the same
between perturbative and numerical results, but slopes are steeper for
numerical results.

At $-C_1=0.331$, the increase in $\eta$ due to the
plaquette term and its suppression by the rectangular loops 
are almost in balance in a wide range of $\beta$, $2.5 \le \beta$, 
and $\eta$ stays close to one. 
However, the detailed behavior of these effects depends on $\beta$ and $\xi_R$. 
$\eta_2$ and $\eta_3$ decrease from $\beta=2.5$, and 
exhibit  shallow dips at $\beta \sim 4.5$ and $\beta \sim 6.0$,
respectively.
These behaviors are qualitatively different from the
prediction of one-loop perturbative calculation.

At $-C_1=1.4088$, the effects of the rectangular loop become stronger 
than those of the plaquette term,
and $\eta$ becomes less than one. 
It is almost independent of $\beta$ in the range 
$1.0 \le \beta \le 2.5$.  This is again qualitatively different from the
perturbative result.

As $\xi_R$ increases, the effects of the two terms in the action are
more enhanced, and the balance becomes more subtle.

In the continuum limit, $\eta$ approaches unity. 
Then Iwasaki action is close to the continuum limit in the region 
$2.5 \le \beta$ for $\xi_R=2$ and 3.
Particularly around $\beta \sim 2.5$, $\eta$ is close to unity.
This means that the calibration of
$\eta$ is not necessary until high precision simulation is carried out.

In the case of DBW2 action, $\eta$ is not close to one. 
Then, as far as $\eta$ is concerned, 
it is not close to the continuum limit in this $\beta$ region.
However, $\eta$ is almost independent of $\beta$.
This is a good property for the simulation of physical quantities 
on anisotropic lattices, 
because the rough calibration of $\eta$ is sufficient to determine $\xi_B$
in this action.

For Symanzik action, the deviation of $\eta$ from unity is remedied
compared with standard action, but the slope of $\eta$ is steep,
and it becomes $\sim 10\%$ at around $\beta \sim 5.0$. 
Therefore detailed calibration of $\eta$ is necessary.

For the $\beta$ and $\xi$ ranges that we have studied, 
the differences between $\eta$ and $\eta^{Nosub}$ are small 
for all the improved actions.
For Symanzik and Iwasaki actions, the difference is $\sim 1\%$, 
and for DBW2 action, it is a few $\%$. 
Therefore it is safe to use $\eta^{Nosub}$ except 
in the case of very precise simulations. 
This is good news, because the calculation of $\eta$ requires much 
CPU time.


Further data on $\eta$ for larger $\xi_R$ and smaller
$\beta$ will be reported in the
forthcoming publications, because
the calculation of $\eta$ at smaller $\beta$ and larger $\xi_R$ 
requires much more CPU time.

\noindent
ACKNOWLEDGMENTS

This work has been done with SX-5 at RCNP and VX-4 at
Yamagata University. We are grateful for the members of RCNP for kind
supports.\\

\appendix
\section{Optimal Radius of Integration For Iwasaki and Symanzik Actions}
If $R$ is an external source field for link variable $U$, 
the link integration of $U$ is given by
\begin{equation}
 \langle U \rangle=    \frac{1}{Z} \frac{dZ(R)}{dR^{\dagger}}
 =\frac{\int D[U] U exp(Tr(RU^{\dagger}+U R^{\dagger}))}
      {\int D[U] exp(Tr(RU^{\dagger}+U R^{\dagger}))} \\
\label{lint1}
\end{equation}
where $Z(R)$ is expressed by the modified Bessel function
$I_{1}$ \cite{brower,forcrand}.
\begin{equation}
 Z(R)=\oint \frac{dx}{2 \pi i} e^{xQ} \frac{1}{z} I_{1}(2z)
\label{zr}
\end{equation}
and
$$ z=\left(\frac{P(x)}{x}\right)^{\frac{1}{2}}, $$
$$ Q=2Re(\det(R)),$$
\begin{equation}
\begin{array}{ll}
  P(x)= 1+xTr(R R^{\dagger}) \\
\hspace*{1cm} +\frac{1}{2} x^2 \left[(Tr(R R^{\dagger}))^2 
              - Tr((R R^{\dagger})^2)\right]
               +x^3 \det(R R^{\dagger}). 
\end{array}
\end{equation}
Similarly $dZ(R)/dR$ is
written by the modified Bessel function $I_{1}$ and $I_{2}$
\cite{brower,forcrand}.

\begin{equation}
 \frac{dZ(R)}{dR^{\dagger}}
 =\oint \frac{dx}{2 \pi i} x e^{xQ} \frac{1}{z}
   I_{1}(2z) \frac{\partial Q}{\partial R^{\dagger}} \\
 + \oint \frac{dx}{2 \pi i} \frac{e^{xQ}}{P(x)} I_{2}(2z)
    \frac{\partial P(x)}{\partial R^{\dagger}}
\label{dzdr}
\end{equation}   
The path of the integration is a closed circle on the complex
plane $x$. In principle it is arbitrary, but numerical
integration requires adequate radius. In the case of the standard
action, the adequate radius was studied by
Scheideler \cite{scheid}.

The arguments of the modified Bessel functions become rather large
and we apply asymptotic expansion.
In this article we use Simpson method for the
numerical integration, and search for the region of $r$ and the number of
the division $N$ where $\langle U \rangle$ is stable against the change of $r$.

\begin{figure}
\begin{center}
\scalebox{0.60}{ { \includegraphics{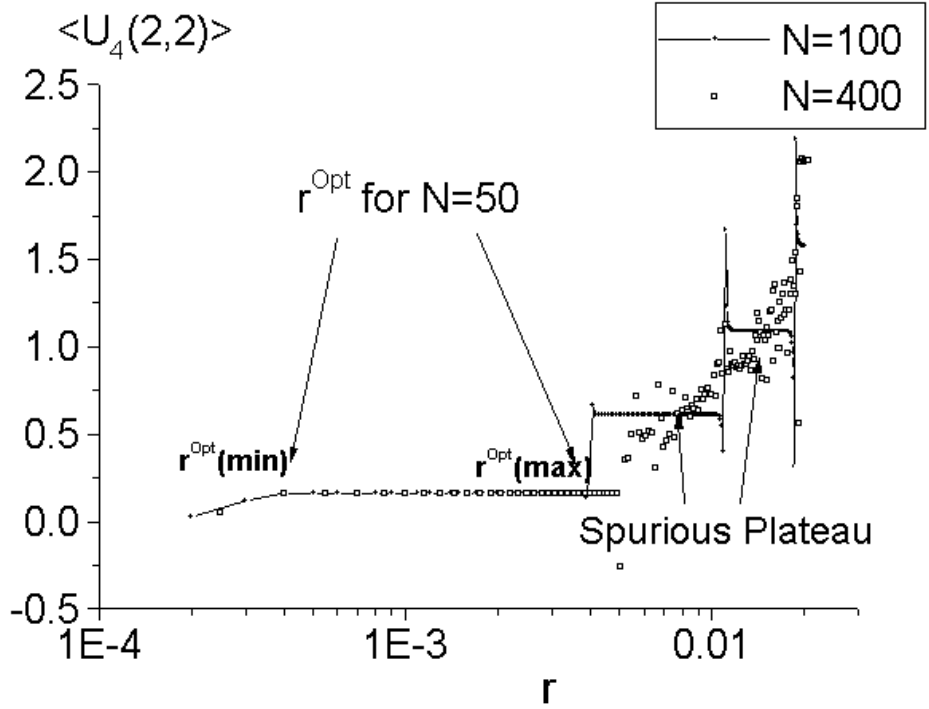} } }
\caption{Radius dependence of $\langle U_4(2,2) \rangle$. The integration is 
carried out on a configuration of Iwasaki action, fully thermalized at
$\beta=3.05$ and $\xi_B=2.0$. An integrated link is located at the
center of the lattice, directing in the 4-th direction. We have shown
the (2,2) elements of the unitary matrix $U$.}   
\label{radius_1}
\end{center}
\end{figure}

An example of the $r$ dependence of a $\langle U \rangle$ is shown in 
Fig.\ref{radius_1}.
It is found that when the number of the devision is $N=100$, there appears
some spurious plateaus, which disappear when $N=400$. However there is a
region of $r$ where $\langle U \rangle$ is stable 
in the change of $N$, which is
the optimal region of integration for $N=100$. The optimal region
increases a little when $N=400$. 
In this article we choose $N=100$ and proceed to determine the optimal
region of $r$ ($r^{opt}$).

These plateaus shown in Fig. \ref{radius_1} are observed when
Taylor expansions of the modified Bessel functions are applied. 
Then they are due to the
difficulty in the numerical integrations given by 
Eq.\ref{zr} and Eq.\ref{dzdr}.
Therefore it is important to find the optimal $r$ region.

\begin{figure}
\begin{center}
\scalebox{0.55}{ { \includegraphics{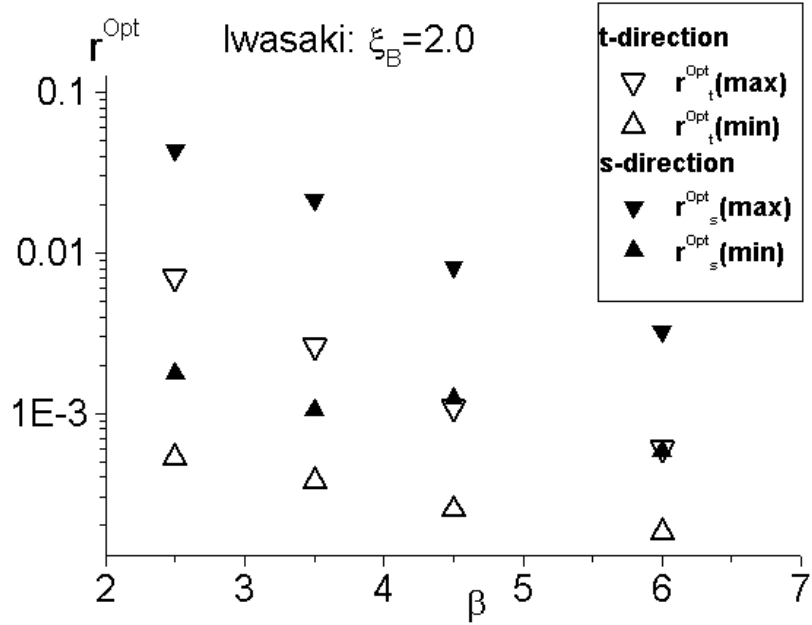} } }
\caption{Examples of the difference between $r^{opt}_{space}$ and
 $r^{opt}_{time}$ for Iwasaki action at $\xi_{B}=2.0$.
 }
\label{opt-iwa}
\end{center}
\end{figure}

For some set of $\beta$ and $\xi_B$, we have 
obtained the minimum and maximum of $r^{opt}$ as shown in the
Fig. \ref{radius_1}, for space-like and time-like links separately. 
When $\xi > 1.0$, $r^{opt}$ of time-like links
($r^{opt}_{t}$) is smaller
than that of space-like links ($r^{opt}_{s}$). 
Examples of the difference is shown in the
Fig. \ref{opt-iwa}.
It seems that the difference becomes larger as $\beta$ and $\xi_B$
increase. The
similar properties are observed in the case of standard action.

\begin{figure}
\begin{center}
\scalebox{0.78}{ { \includegraphics{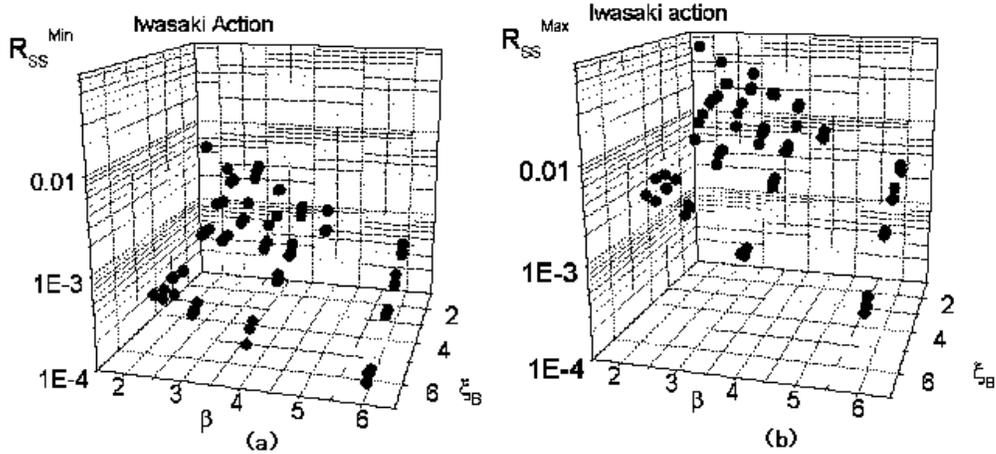} } }
\caption{A compilation of $R_{min}$, (a) and $R_{max}$ (b) of Iwasaki 
action 
in the range $2.0 \le \beta \le 6.0$ and $2.0 \le \xi_{B} \le 6.0$. }
\label{rssiwa_all}
\end{center}
\end{figure}

We proceed to the parameterization of the $r^{opt}(\beta,\xi)$.
The $\beta$ and $\xi$ ranges are $2.0 \le \beta \le 6.0$, $1.8 \le \xi_B
\le 6.5$ for
Iwasaki action and $4.5 \le \beta \le 8.0$, $1.7 \le \xi_B \le 5.8$ for
Symanzik action.
The $r^{opt}_{space}(min)$ and $r^{opt}_{space}(max)$ are shown in the 
Fig. \ref{rssiwa_all}. 
%
%
They decrease with $\beta$ and $\xi$ and seems to be parametrized as 
\begin{equation}
  r^{opt} = a * \exp(-b \beta - c \xi_B)
\label{r-parameter}
\end{equation}
Then we define $r^{opt}(mid)$ as,
\begin{equation}
  \log(r^{opt}(mid)) = (\log(r^{opt}(min)) + \log(r^{opt}(max)))/2.
\end{equation}
and then fit them by Eq. \ref{r-parameter}. 
 For $\log(r^{opt})$, it becomes a multiple regression fitting. The
coefficients are determined by the least square method.  The results for
Symanzik and Iwasaki actions are summarized in the table
\ref{r-fit}.
\renewcommand{\arraystretch}{1.0}
\begin{table}[h]
\begin{center}
\caption{ The multiple regression fit of $r^{opt}(mid)$. The 60 data
and 29 data points are used to determine the coefficients $a$, $b$, $c$
for Iwasaki and Symanzik actions respectively.}
\label{r-fit}
     \vspace {0.5cm}

\begin{tabular}{|c|c|c|c|c|c|c|}
     \hline
     \multicolumn{1}{|c|}{Action} &
     \multicolumn{1}{|c|}{ } &
     \multicolumn{1}{|c|}{ a}&
     \multicolumn{1}{|c|}{ b}&
     \multicolumn{1}{|c|}{ c}\\
     \hline
      $Symanzik$  &$r^{opt}_{s}(mid)$  &0.5563 &0.5479 &0.5336 \\
      $(26 data)$ &$r^{opt}_{t}(mid)$  &0.06244 &0.4213 &0.6568 \\
     \hline
      $Iwasaki$   &$r^{opt}_{s}(mid)$  &0.08663 &0.5507 &0.4315 \\
      $(61 data)$ &$r^{opt}_{t}(mid)$   &0.01682 &0.5139 &0.5261\\
     \hline
\end{tabular}
\end{center}
\vspace{0.5cm}
\end{table}  
We have checked that $r^{opt}(mid)$ with the parameter given by
table \ref{r-fit} is located
between $r^{opt}(max)$ and $r^{opt}(min)$; namely it stays within the
optimal radius of integration through out the data points.

$r^{opt}$ region may change with the position of link on a configuration 
and also with configurations.
The results shown in Fig.\ref{rssiwa_all} are obtained for
a link at the center of the configuration in space and time
directions, which are fully thermalized.
However the fluctuation of the $r^{opt}$ region is 
not large. If we choose, $r^{opt}(mid)$,
it has been in an optimal region of $r$ for all link variables and 
configurations.

Let us proceed to discuss the effects of the link integration method.
In the case of improved actions, the number of link $U$ which are 
simultaneously
integrated in a Wilson loop becomes much smaller than the case of standard
action, because in the case of improved action the background fields
$R$ of Eq. \ref{zr} extend wider range due to the 6-link
rectangular loop in the action.
Therefore the effect of link integration method is reduced in these
cases, and is not effective for the
calculation of smaller Wilson loops.

\begin{figure}
\begin{center}
\scalebox{0.65}{ { \includegraphics{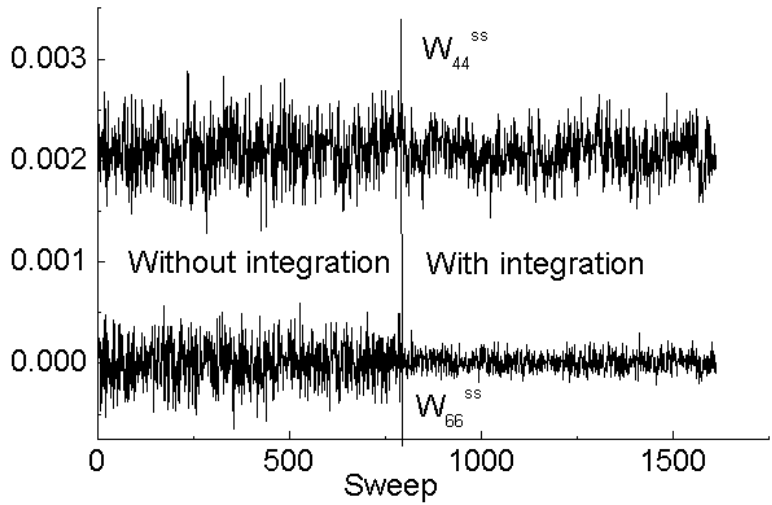} } }
\caption{An example of the  suppression of the fluctuation of Wilson loops 
for Iwasaki
action at $\beta=2.5$ and $\xi_B=2.0$ }
\label{lint-w}
\end{center}
\end{figure}

The example of the suppression of the fluctuation is shown in the
Fig. \ref{lint-w}.
%
%
The suppression of the
fluctuation is impressive for $W(6,6)$ but not for $W(4,4)$.
The similar properties are observed for the Symanzik
action of $W(8,8)$ and $W(4,4)$ at $\eta=4.5$. 
The link integration needs much
CPU-time, and therefore the link integration method is effective for the
calculation of large Wilson loops in the confined phase or very close
to the transition point.


\begin{thebibliography}{99}
\bibitem{karsch} F.~Karsch, Nucl. Phys. B205[FS5] (1982), 285.
\bibitem{klassen} T.~R.~Klassen, Nucl. Phys. B533 (1998) 557.
\bibitem{eta_bielefeld} J.~Engels, F.~Karsch and T.~Scheideler,
Nucl. Phys. B564 (2000) 303.
\bibitem{sakai} S.~Sakai, T.~Saito, and A.~Nakamura,
Nucl. Phys. B584 (2000) 528.
\bibitem{iwasaki} Y.~Iwasaki, Nucl. Phys. B258 (1985), 141; Univ. of Tsukuba
preprint UTHEP-118 (1983).
\bibitem{symanzik} K.~Symanzik, Nucl. Phys. B226 (1987) 187\\
 M.~Leuscher and P.~Weise, Phys. Lett. B158 (1985) 250. 
\bibitem{dbw2} QCD-TARO Collaboration Ph. de Forcrand et al.,
Nucl. Phys.  B577 (2000) 263 (hep-lat/9911033).
\bibitem{parisi}  G.~Parisi, R.~Petronzio and F.~Rapuano, 
Phys. Lett. 128B (1983) 418, Nucl. Phys. B205[FS5] (1982) 337.
\bibitem{brower}  R.~Brower, P.~Rossi and C.~I.~Tan, Nucl. Phys. B190[FS3] 
(1981) 699.
\bibitem{forcrand}  Ph. de~Forcrand and C.~Roiesnel,
Phys. Lett. B31 (1985) 77.
\bibitem{yoshie}  T.~Yoshie, Nucl. Phys. B(Proc Suppl) 63 (1998) 3.   
\bibitem{qcdtaro_compilation}  QCDTARO, private communication.
\bibitem{cella}  G.~Cella, G.~Curci, A.~Vicere and B.~Vigna, 
Phys.  Lett. B333 (1994) 457.
\bibitem{kaneko}  Y.~Iwasaki, K.~Kanaya, T.~Kanenko and T.~Yoshie,
Nucl. Phys. B(Proc Suppl) 53 (1997) 429.   
\bibitem{sakai96}  A.~Nakamura, T.~Saito and S.~Sakai
Nucl. Phys. B(Proc Suppl) 63 (1998) 424.
\bibitem{lambda}  Y.~Iwasaki and S.~Sakai, Nucl. Phys. B248 (1984) 441.
\bibitem{nakamura} G.~Burgers, F.~Karsch, A.~Nakamura and I.~O.~Stamtescu,
Nucl. Phys. B304 (1988) 587.
\bibitem{woch} F.~R.~Brown and T.~J.~Woch,
Phys. Rev. Letters 58 (1987) 2394.
\bibitem{creutz2} M.~Creutz, Phys. Rev. D36 (1987) 515.
Phys. Rev. Letters 58 (1987) 2394.
\bibitem{sakai1} S.~Sakai, A.~Nakamura and T.~Saito,
Nucl. Phys. A638 (1998) 535c.
\bibitem{sakai2} S.~Sakai, A.~Nakamura and T.~Saito, 
Nucl. Phys. B(Proc Suppl) 83 (2000) 399.
\bibitem{petronzio}  G.~Martinelli, G.~Parisi and R.~Petronzio, 
Phys. Lett. 100B (1981) 435.
\bibitem{lepage}  G.~P.~Lepage and P.~B.~Mackenzie,
Phys. Rev. D48 (1993) 2250.
\bibitem{scheid} T.~Scheideler, Dr. thesis, Bielefeld(1997).
\end{thebibliography}
\end{document}